\documentstyle[12pt]{article}

\begin{document}

\thispagestyle {empty}
\rightline{TIFR/TH/94-24}
\rightline{IC/94/147}
\rightline{hep-th 9407015}
\vskip 1.5 cm

\begin{center}

{\large\bf BOUNDARY DYNAMICS IN DILATON GRAVITY}\break
\end{center}
\vskip .5 cm
\begin{center}
{\bf Sumit R. Das}

{\sl Tata Institute of Fundamental Research \break
Homi Bhabha Road, Bombay 400005, INDIA \break
e-mail: das@theory.tifr.res.in}\break

and
\vskip .3 cm
{\bf Sudipta Mukherji}

{\sl International Centre for Theoretical Physics \break
I-34100 Trieste, ITALY\break
e-mail: mukherji@ictp.trieste.it}\break
\end{center}
\vskip .5 cm
\begin{center}
{\bf ABSTRACT}
\end{center}
\vskip .2 cm
We study the dynamics of the boundary in two dimensional dilaton
gravity coupled to $N$ massless scalars. We rederive the boundary
conditions of \cite{DAM1} and \cite{CHVER} in a way which makes
the requirement of reparametrization invariance
and the role of conformal anomaly explicit. We then
study the semiclassical behaviour of the boundary in the $N=24$ theory
in the presence of
an incoming matter wave with a constant energy flux spread over a
finite interval. There is a critical value of the matter energy
density below which the boundary is stable and all the matter is
reflected back. For energy densities greater than this critical value
there is a similar behaviour for small values of the {\em total}
energy thrown
in. However, when the total energy exceeds another
critical value the boundary
exhibits a runaway behaviour and the spacetime develops singularities
and horizons.
\newpage
\def\ben{\begin{equation}}
\def\een{\end{equation}}
\def\bea{\begin{eqnarray}}
\def\eea{\end{eqnarray}}
\def\p{\partial}
\def\xp{X^{+}}
\def\xm{X^{-}}
\def\yp{Y^+}
\def\ym{Y^-}
\def\half{{1 \over 2}}
\def\pu{\partial_u}
\def\pv{\partial_v}
\def\rh{{\hat \rho}}
\def\kh{{\hat \kappa}}
\def\klog{{\rm log}}
\def\pp{\partial_+}
\def\ppm{\partial_-}
\def\hb{h_B}
\def\fb{f_B}
\def\gb{g_B}
\def\qb{q_B}
\def\kk{{{\hat \kappa}\over\kappa}}
\def\ep{\epsilon}
\def\eplus{{\rm log}(1+\ep)}
\def\eminus{{\rm log}(1-\ep)}
\def\aep{\epsilon^{(1)}}
\def\bep{\epsilon^{(2)}}
\def\sqd{{\sqrt{\Delta}}}
\def\sqs{{\sqrt{\Sigma}}}
\def\sups{{\sqrt{\Upsilon}}}
\def\sqratio{{\sqrt{\sqd/\sqs}}}
\def\ktan{{\rm tan}}

\def\prd{{\em Physical Review~  }}
\def\prl{{\em Physical Review Letters~}}
\def\pl{{\em Physics Letters~}}
\def\np{{\em Nuclear Physics}}

\section{Introduction}

In a recent paper \cite{DAM1},
we have studied a model of two dimensional gravity
and massless matter field with a boundary in space time
\cite{SVV,CHVER}. The reason for
introducing such a boundary in the model is many-fold.
In the two dimensional reduction of four dimensional spherically
symmetric space time such a boundary may be thought as
the origin of the radial coordinate.
Moreover semiclassical studies of dilaton gravity models in two
dimensions inevitably lead to a boundary beyond which the theory
becomes strongly coupled. In a ficucial coordinate system $(u, v)$ the
boundary may be chosen to be the line $u = v$.  We will refer to this
coordinate system as the ``fixed boundary gauge''.  In these
coordinates the conformal factor of the metric and the dilaton are
independent fields. However, $u$ and $v$ are not asymptotically
minkowskian.  Alternatively we can work in Kruskal coordinates which
are simply related to the asymptotically minkowskian coordinates. In
these coordinates the dilaton is equal to the liouville mode and the
boundary is {\em dynamical}.  Its world line is determined by the
matter energy momentum tensor, through a {\em boundary equation}
derived in \cite{DAM1,CHVER}. In the fixed boundary gauge the
same equation determines the relationship between the dilaton and the
liouville mode.

In the full quantum theory the boundary undergoes quantum fluctuations
which are determined by the boundary equation, which has to be now
regarded as an operator equation. (In the fixed boundary gauge, these are
the dilaton and liouville fields which fluctuate.) The dilaton-gravity
sector of the theory has no physical propagating degrees of freedom,
but the degree of freedom corresponding to the boundary survives. Thus
we get an idea about the strength of quantum gravity in such simple
model by studying boundary fluctuations, as done in \cite{DAM1}.

In some sense the boundary acts as a model for the high energy, strong
coupling physics of the problem. Just as in the Callan-Rubakov effect
we would like to model the high energy physics by suitable boundary
conditions. The boundary conditions must respect some minimum
requirements, like conservation of charges which should be conserved.
In general such minimal requirements may not be restrictive enough to
enable us to get insight into the real problem. However, it is also
common that minimal requirements turn out to be rather restrictive and
allow one to derive ``model indpendent'' consequences. The
boundary conditions which arise in the
Callan-Rubakov effect is such an example \cite{DCAL}.

In the present problem, the boundary conditions must respect
conservation of energy momentum and the general covariance of the
underlying theory.  In a gauge fixed theory the latter means that the
reparametrization constraints must be obeyed (in a conformal gauge
this is the same as maintaining conformal invariance).

The most general boundary conditions which respect these requirements
were derived in \cite{CHVER,DAM1}. The conditions of general
covariance and energy-momentum conservation turn out to be rather
restrictive. Assuming that the matter is reflected off the boundary
separately, the boundary conditions in the dilaton gravity sector is
characterized by a single parameter $\beta$,
which is related to the value of the dilaton field at the boundary.
In \cite{DAM1} a
``boundary equation'' was derived starting from the fixed boundary
gauge. The behavior of
the boundary and the resulting space-time structure is similar for all
non-zero values of $\beta$, while $\beta =0$ (which are the boundary
conditions of \cite{RST,SVV}) is a rather singular limit.
In fact the boundary conditions arise dynamically when one adds
a boundary cosmological constant term proportional to $\beta$
to the two dimensional action
\cite{STHOR,STROM}, which immediately explains
why $\beta = 0$ is rather special. Our model is in fact similar to
the boundary conformal field theories considered recently in \cite{BCON}.

In \cite{DAM1} the semiclassical boundary equation was
solved in the $N=24$ theory for an incident shock wave.
For $\beta \ne 0$ there is a critical value of the strength of the
shock wave, i.e. the {\em total} energy. Below this critical value the
boundary has a stable timelike evolution and no black hole is formed. When
the energy exceeds the critical value the boundary runs away with ever
increasing acceleration, approaching the speed of light. At the same
time a black hole is formed which subsequently evaporates.  The
stability of the boundary for small disturbances have been recently
analyzed in \cite{STHOR} for arbitrary $N$ and an intriguing
proposal based on the similarity of the model with open string theory
has been put forward for the $N = \infty$ theory in \cite{STROM}.

Note that a limit on the strength of the shock wave does not set any
limit on the {\em energy density} for an aribitrary pulse.
The $\beta = 0$ theory, which is
essentially the RST model, can be solved for arbitrary distribution of
incoming matter energy and a black hole is formed when the {\em energy
density} exceeds a critical value. However, in this model the boundary
becomes space-like whenever this critical value is exceeded.  It is of
interest to know the behavior of the $\beta \ne 0$ model
for more general incoming matter
distribution and ask whether there is some critical value for the {\em
energy density} as well.

In this letter we first rederive the boundary condition and the
boundary equation for general $N$ in the Kruskal gauge. This
derivation makes the role of reparametrization invariance and the
anomaly in the boundary conditions explicit. We then solve the
semiclassical boundary equation for $N = 24$ for an
incoming matter distribution which consists of a constant energy
density in a finite interval of retarded time. We find that for the
boundary to become unstable and a black hole to form, there is a
critical value of the energy density as well as the total energy.

\section{The Boundary Equation}

The classical action of the model is given by
\ben
S = {1\over{2\pi}}\int du dv {\sqrt g}[e^{-2 \phi}(R - 4 (\nabla \phi )^2
- 4 \lambda^2 + {1\over 2} (\nabla f)^2 ]\label{eq: one}
\een
where $g_{ab}(u, v)$, $\phi (u, v)$ and $f^{i}(u, v)$ are the two
dimensional metric, dilaton and the matter fields. $R$ is the
curvature scalar and $\lambda$ is the comological constant which we
set to one by properly choosing the scale.

We will use the conformal gauge where $g_{ab} = e^{2\rho} \eta_{ab}$.
Since the matter is conformally coupled its quantum effects are
entirely contained in the Weyl anomaly which means that there is a
liouville term in the action. There are additional
liouville terms coming from the measures of integration of the
conformal mode $\rho$, the dilaton and the ghosts. These measures are
chosen so as to ensure that the Hawking radiation consists of only
physical propagating particles, which are the matter fields $f^i$.
Following RST one can use the freedom to add a local counterterm to
make the semiclassical model solvable. The resulting action may be
written as
\ben
S = {\kappa \over \pi}\int du~dv[2 \Omega \pu\pv \rh - e^{2\rh} - 2(1 - {1
\over \kappa})\pu \rh \pv
\rh + \half \pu f^i \pv f^i + ({\rm ghosts})]
\label {eq:two}
\een
where we have defined the fields
\ben
\Omega = {1 \over \kappa}[e^{-2\phi}+\kappa\phi]~~~~\rh=\rho-\phi
\label{eq:three}
\een
and $\kappa = {N \over 24}$.

It is also possible to write down models which are exactly conformally
invariant \cite{BCAL,DEAL}.  These models have an action
which is identical to (\ref{eq:two}). However the expressions for
$\Omega$ and $\rh$ in terms of $\phi$ and $\rho$
are different. The boundary conditions and
the boundary equation which we will derive are valid for these
theories as well. However the interpretation of the solution in terms
of the space-time structure would be different.

The equations of motion following from (\ref{eq:two}) are very
simple
\ben
\pu\pv\rh = \pu\pv f^i = 0 ~~~~~~\pu\pv\Omega = -e^{2\rh}
\label{eq:four}
\een
Following \cite{SVV} we introduce two chiral fields $\xp(u)$ and
$\xm(v)$ and solve the equation for $\rh$
\ben
e^{2(\rho -\phi)} = \p_u \xp (u) \p_v \xm (v)\label{eq:five}
\een
The equation for $\Omega$ has the general solution
\ben
\Omega = - \xp (u) \xm (v) + g^+(u) + g^-(v) + K\label{eq:six}
\een
where $K$ is a constant and $g^\pm$ are arbitrary chiral functions.
We can trade these functions for two other chiral fields
$\yp (v)$ and $\ym (u)$ which are defined as
\ben
\p_u g^+(u) = Y^-(u)\p_u\xp (u),~~~\p_v g^-(v) = \yp (v) \p_v \xm (v),
\label{eq:seven}
\een

In the gauge fixed theory the requirement of general covariance of
the underlying model is imposed by setting the total energy momentum
tensor to be zero
\ben
T^g_{uu}+T^m_{uu}+T^{gh}_{uu}=0~~~~T^g_{vv}+T^m_{vv}+T^{gh}_{vv}=0
\label{eq:eight}
\een
where $T^g, T^m, T^{gh}$ stand for the graviton-dilaton, matter and
the ghost energy momentum tensors respectively. In what follows, the
ghosts completely decouple. The matter part $T^m$ is standard, while
the graviton-dilaton part has the following expressions
\bea
T^g_{uu} & = & \kappa\p_u \ym \p_u \xp + {\kh \over 2}
[(\p_u{\rm log}\p_u\xp)^2
- 2\p_u^2{\rm log}\p_u \xp] \nonumber\\
T^g_{vv} & = & \kappa\p_v \yp \p_v \xm + {\kh \over 2}
[(\p_v{\rm log}\p_v \xm)^2
- 2\p_v^2{\rm log}\p_v\xm]\label{eq:nine}
\eea
where $\kh = {N - 24\over{24}}$.
Note that in (\ref{eq:eight}) above, the matter part $T^m$ contains
a vacuum energy term coming from normal ordering effects. Alternatively
in the semiclassical theory the vacuum has to be defined in terms
of modes which are positive frequency with respect to asymptotically
minkowskian coordinates and there is a term which comes from the
anomalous transformation of the energy momentum tensor to some other
globally valid coordinate system.

We now derive the general boundary condition. The minimal
requirements which the theory with these boundary conditions must
satisfy are

\begin{enumerate}

\item The dilaton field must be a constant along the
boundary

\item The matter must be perfectly reflected from the
boundary

\item The reparametrization constraints (\ref{eq:eight})
must be obeyed.

\end{enumerate}

The last condition guarantees that the boundary
conditions are consistent with the general covariance of the
model.

Consider a timelike boundary in the theory specified by the
equation
\ben
v = f_B (u)
\label{eq:ten}
\een
or its inverse equation
\ben
u = g_B (v)
\label{eq:eleven}
\een
The condition (2) above and the constraint (\ref{eq:eight}) imply that
the total energy momentum tensor must be perfectly reflected.
This means that the right moving energy momentum tensor
must be set to be equal to the left moving
energy momentum tensor after performing a coordinate transformation
by the function (\ref{eq:eleven}). In doing so one must remember that
energy momentum tensors transform anomalously with a Schwarzian
derivative term.

We will impose Dirichlet boundary conditions on the matter fields
\ben
f^i(g_B(v),v)=0
\label{eq:twelve}
\een
We will also impose similar boundary conditions on the ghosts. This
means that general covariance requires that the gravity energy
momentum tensor must obey
\ben
T^g_{vv}(v) = (\pv g_B(v))^2~T^g_{uu}[g_B(v)] + \kh \{ g_B(v),v \}
\label{eq:thirteen}
\een
where $\{ .,. \}$ denotes a Schwarzian derivative
\ben
\{ f,x \} = {f''' \over f'}- {3 \over 2}({f'' \over f'})^2
\een

In the bulk theory independent reparametrizations of $u$ and $v$ are
still symmetries in the conformal gauge. This is no longer
true if we choose the boundary to be a fixed line $u = v$ as in
the fixed boundary gauge. However, {\em one} of the conformal
reparametrizations still remain a symmetry and we can use this to
choose $\xp (u) = u$. This leaves $\xm (v)$ as the dynamical
degreee of freedom, which is determined by the constraints.
This is the gauge used in \cite{SVV,DAM1}. In the
semiclassical theory, the constraints determine the function
$\xm (v)$ in terms of the energy momentum tensor of the matter.
In the quantum theory the fluctuations of $\xm (v)$ are determined
in terms of the matter fluctuations via the operator constraints.
Since in this gauge $g_B(v) = 1$ the Schwarzian derivative term
vanishes. However the nontriviality of the transformation properties
of the energy momentum tenaor are contained in the expression for
$T^g_{vv}$ in (\ref{eq:nine}).

Alternatively one may fix the Kruskal gauge $\xp (u) = u$ and
$\xm (v) = v$. The boundary cannot be fixed to be some predetermined
line any more. Rather the boundary curve is now dynamically determined,
as we shall see soon. In the following we will use the Kruskal gauge
and accordingly replace the indices $u$ and $v$ by $+$ and $-$ respectively.

The condition that the dilaton is constant along the boundary follows
from the solution for $\Omega$ in (\ref{eq:six}) and (\ref{eq:seven})
\ben
(\ym(\xp)-\xm)+(\pp f_B)(\yp(\xm)-\xp) \vert_{\xm=f_B(\xp)}=0
\label{eq:fourteen}
\een
Let us introduce a new quantity
\ben
h_B(\xp) = {\sqrt{\pp f_B(\xp)}}
\label{eq:fifteen}
\een
The condition (\ref{eq:fourteen}) may be satisfied by setting
\bea
\yp(\xm) & = & g_B(\xm) + F[h_B(g_B(\xm)),\xm] \nonumber\\
\ym(\xp) & = & f_B(\xp) - {F[h_B(\xp),\xp] \over h_B^2}
\label{eq:sixteen}
\eea
where $F[h_B(x),x]$ is some general functional of $h_B(x)$ and
a function of $x$.

We now use these expressions for $Y^\pm$ to evaluate $T^g_{++},T^g_{--}$
in (\ref{eq:thirteen}) and obtain a functional differential equation for
$F$. In the Kruskal gauge one simply has $T^g_{++}=\pp \yp$ etc. However
since the boundary curve is nontrivial, the term in
(\ref{eq:thirteen}) which involves $\kh$ is nonvanishing. One gets the equation
\ben
\pp F + \pp h_B {\delta F \over \delta h_B} = (\pp~ \klog~h_B)F
+{\kh \over \kappa}[(\pp~\klog~h_B)^2 - \pp^2~\klog~h_B]
\label{eq:seventeen}
\een
This has the most general solution
\ben
F[h_B(\xp),\xp] = \beta h_B(\xp) + {\kh \over \kappa}\pp~\klog~h_B
\label{eq:eighteen}
\een
where $\beta$ is an arbitrary parameter. This completes the derivation
of the most general boundary condition which satsifies the requirements
stated above. The term involving $\kh$ is clearly the result of the
anomalous transformation law of the energy momentum tensor.

Finally, using the above expression for $F$ one may derive the equation
of motion of the boundary. This is simply the statement that the total
$T_{++}$ vanishes and is given by
\ben
\kappa h_B^2 + \kappa \beta \pp h_B + \kh \pp^2~\klog~h_B
+T^m_{++}-{\kappa \over 2 (\xp)^2} = 0
\label{eq:nineteen}
\een
The last term  is the standard vacuum energy which appears in the Kruskal
gauge.

So far we have not specified the value of the dilaton field on the boundary.
At this stage this may appear as a second parameter specifying the boundary.
However, as we shall se below, this is actually determined by the condition
that the vacuum solution of the boundary equation corresponds to flat space
with a linear dilaton.

The boundary equation (\ref{eq:nineteen}) determines the shape of the
boundary curve in terms of the matter energy momentum tensor. Once this is
known, we can use the expressions for $\ym,\yp$ to write the solution
for $\Omega$ and hence derive the space-time structure. In the following
we will concentrate on the behaviour of the boundary itself.

It is useful to rewrite this equation in terms of the quantity
\ben
\qb (\xp) = \xp \hb (\xp)
\label{eq:twenty}
\een
In terms of $\qb$ the equation (\ref{eq:nineteen}) becomes
\bea
\kk (\xp\pp)^2~\klog~\qb & + (\beta\qb - \kk)(\xp\pp)\klog~\qb &
\nonumber\\
&+(\qb^2 - \beta\qb + \kk - \half)& = - {(\xp)^2 \over \kappa}T_{++}^M
\label{eq:twoone}
\eea
As noted in \cite{STHOR} this represents the equation of motion
for a particle driven with some external force and subjected to some
damping. In vacuum $T^M_{++}=0$ and the external force is the
gradient of a potential. Consequently one may expect that the
ground state is characterized by a constant $\gb$ which is the
minimum of this potential. However, for this to occur for real $\gb$
one must have $\beta^2 + 2 > 4\kk$. For general $N$ one cannot, therefore,
choose $\beta = 0$ \cite{CHVER}.
Furthermore for the vacuum solution to be stable under small deviations,
one must have $\kh \geq 0$ and $\beta > {{\sqrt{2}}\kh \over \kappa}$.

In the rest of the paper we shall limit ourselves to $N=24$ for which
the vacuum is perturbatively stable.

\section{Semiclassical Solution with a finite duration wave}

We now investigate the stability of the boundary for arbitrary incoming
energy and energy density for $N = 24$. Even in this case,
we have not been able to solve the equation (\ref{eq:twoone}) for
arbitrary $T_{++}$. However to understand the nature of threshold of
instability and black hole formation we only need to have a situation
where we can tune the energy density and the total energy independently.
We will thus solve the above equation for an incoming matter wave
which has a constant energy density for a finite interval of retarded
time. Thus we have a matter energy momentum tensor of the form
\ben
T^M_{++}={\alpha \over 2 (\xp)^2}[\theta(\xp - 1 + \ep)-\theta(\xp - 1 - \ep)]
\label{eq:twotwo}
\een
which corresponds to a constant energy density ${\alpha \over 2}$ in the
interval $1-\ep < \xp < 1 + \ep$. The total energy is thus $M = \alpha\ep$.

In the following we will name the various regions of the $\xp$ space
as follows (i) Region I : $\xp < 1 - \ep$ (ii) Region II $1-\ep < \xp <
1 + \ep$ (iii) Region III $\xp > 1 + \ep$.

In Regions I and II the solution is given by
\ben
\qb (\xp) = \half[\beta + \sqd{(\xp)^{\delta} + d \over (\xp)^\delta - d}]
\label{eq:twothree}
\een
where $d$ is an integration constant. In (\ref{eq:twothree}) we have
made the following definitions
\ben
\Delta = \beta^2 + 2~~~~~~~~~\delta = {\sqd \over \beta}
\label{eq:twofour}
\een

We want to have initial conditions
such that the space time is in the vacuum, and the boundary has a vacuum
hyperbolic solution discussed above, before any matter has come in. This
means that in region I we have to choose $d = 0$. This determines the
value of the dilaton field on the boundary to be the quantity \cite{DAM1}
\ben
e^{-2\phi} \vert_{boundary} = \half[1+\beta(\beta + \sqd)]
\label{eq:twofoura}
\een
Thus the value of the dilaton field is also determined in terms of
the parameter $\beta$ as mentioned above.

The nature of the solution in region II is qualitatively different for
small and large $\alpha$ regardless of the value of $\ep$. Let us define
$\alpha_0 = {\Delta \over 2}$.

\subsection{$\alpha < \alpha_0$}

For $\alpha < \alpha_0$ the solution in
region II is given by
\ben
\qb = \half[\beta + \sups {(\xp)^{\eta}+d_{II} \over (\xp)^{\eta}-d_{II}}]
\label{eq:twofive}
\een
where we have defined
\ben
\Upsilon = \Delta - 2\alpha~~~~~~~~~\eta = {\sups \over \beta}
\label{eq:twosix}
\een
The constant of integration $d_{II}$ is determined by requiring that
$\qb$ is continuous across $\xp = 1- \ep$. This gives
\ben
d_{II} = (1-\ep)^\eta~{\sqd - \sups \over \sqd + \sups}
\label{eq:twoseven}
\een
 From the solution (\ref{eq:twofive}) it is clear that
$\qb$ and hence $\fb$ will become infinite and the boundary will run away
if $(\xp)^{\eta}-d_{II}=0$ for some value of $\xp >
(1 - \ep)$. This requires $d_{II} > (1-\ep)^\eta$. However,
using the definitions of $\Delta$ and $\Upsilon$ it is clear
from (\ref{eq:twoseven}) that $d_{II}$ is always less than
$(1-\ep)^\eta$. Thus there cannot be any runaway of the boundary
in the region II.

The solution in region III is given by (\ref{eq:twothree}) with the
integration constant denoted by $d_{III}$. This is in turn determined
by the solution in region II by matching across $\xp = 1+ \ep$, which
yields the following equation for $d_{III}$
\ben
{(1+\ep)^\delta + d_{III} \over (1+\ep)^\delta - d_{III}}
= {\sups \over \sqd} {(1+\ep)^\eta + d_{II} \over (1+\ep)^\eta - d_{II}}
\label{eq:twoeight}
\een
Since $d_{II} < (1-\ep)^\eta < (1+\ep)^\eta$
it is clear from (\ref{eq:twoeight}) that
$d_{III} < (1+\ep)^\delta$. The solution (\ref{eq:twothree}) then
shows that $\qb$ cannot blow up anywhere in region III as well.

We thus conclude that for $\alpha < {\Delta \over 2}$ the boundary is
stable and remains timelike throughout.

\subsection{$\alpha > \alpha_0$}

For $\alpha > \alpha_0$ the solution in Region II is given by
\ben
\qb = \half[\beta - \sqs~\ktan(\mu~\klog~\xp + c_{II})]
\label{eq:twonine}
\een
where we have defined
\ben
\Sigma = 2\alpha - \Delta ~~~~~~~~\mu = {\sqs \over 2\beta}
\label{eq:thirty}
\een
The intgration constant $c_{II}$ is once again determined by matching
the solution with the vacuum solution at $\xp = 1-\ep$. This gives
\ben
\ktan~c_{II} = {\ktan~\mu\eminus + \gamma \over
\gamma~\ktan~\mu\eminus - 1}
\label{eq:thone}
\een
where we have defined
\ben
\gamma = {\sqrt{\Delta}\over \sqrt{\Sigma}}
\label{eq:thonea}
\een
The solution in region III is still given by (\ref{eq:twothree})
which we rewrite as
\ben
\qb (\xp) = \half[\beta + \sqd{(\xp)^{\delta} + c_{III}
\over (\xp)^\delta - c_{III}}]
\label{eq:twothreea}
\een
$c_{III}$ is determined by matching across $\xp = 1+ \ep$ which
yields
\ben
c_{III} = (1+\ep)^\delta~{(1+\gamma^2)\ktan[\mu~\klog({1+\ep \over 1-\ep})]
\over (1-\gamma^2) \ktan[\mu~\klog({1+\ep \over 1-\ep})] - 2\gamma}
\label{eq:thtwo}
\een

Let us first look for runaways of the boundary in the region III. The solution
shows that this can happen if the value of $c_{III}$ is such that the
denominator in (\ref{eq:twothreea})
vanishes for a value of $\xp > (1 + \ep)$. This means that one must have
$c_{III} > (1+\ep)^\delta$. The solution for $c_{III}$ then shows that
this would require
,for $\gamma \geq 1$,
\ben
-{2\gamma \over \gamma^2 - 1} < \ktan[\mu~\klog({1+\ep \over 1-\ep})]
< - {1 \over \gamma}
\label{eq:thtwoa}
\een
whereas for $\gamma < 1$ one must have either
\ben
\ktan[\mu~\klog({1+\ep \over 1-\ep})] > {2 \gamma \over 1- \gamma^2}
\label{eq:thtwob}
\een
or
\ben
\ktan[\mu~\klog({1+\ep \over 1-\ep})] < -{1 \over \gamma}
\label{eq:thtwoc}
\een

It may be easily checked that in the limiting case of a shock wave this
reproduces the limit on the total energy derived in \cite{DAM1}. This limit
corresponds to $\alpha \rightarrow \infty$ and $\ep \rightarrow 0$ with
the total mass $M = \alpha\ep$ kept fixed. Then (35) reduces
to the condition $M > \beta\sqd$.

For finite values of $\alpha$ and $\ep$ one has a more complex behaviour,
since the various integration constants are periodic functions of $\alpha$ and
$\epsilon$. Let us study the behaviour of the boundary for some given
value of $\alpha > \alpha_0$ and $\beta$ and the extent of the pulse
$\ep$ increasing from zero. It follows from (\ref{eq:thtwoa})-(\ref{eq:thtwoc})
that the conditions for runaway in region III are satisfied
for $\ep$ lying in the range $\ep^{(1)}_n < \ep < \ep^{(2)}_n$ where
\bea
\ep^{(1)}_n & = &{\rm tanh}~ \{ {1 \over 2\mu}[\delta_n\pi
 + \ktan^{-1}({2\gamma \over 1-\gamma^2})] \} \nonumber \\
\ep^{(2)}_n & = & {\rm tanh}~ \{ {1 \over 2\mu}[n\pi
- \ktan^{-1}({1 \over \gamma})] \}
\label{eq:thtwod}
\eea
where $n = 1,2, \cdots$ and
\ben
\delta_n = (n-1) \theta (1 - \gamma) + n \theta (\gamma - 1)
\een
In the above equation the branch of the inverse tangent function has
been chosen to be the interval $[-{\pi \over 2}, {\pi \over 2}]$.
It is easy to check that $\aep_{n} < \bep_n < \aep_{n+1}$

The behaviour of the boundary may be now summarized as follows.
For $\ep < \aep_1$, $\qb$ is finite and the boundary is stable.
For $\aep_1 < \ep < \bep_1$ the boundary runs away in region III.
For $\bep_1 < \ep < \aep_2$ the quantity $\qb$ diverges in region II and
there is no divergence in region III. When $\aep_2 < \ep < \bep_2$ a
new divergence of $\qb$ appears in region III, and the pattern
continues.


In the physical problem a divergence of $\qb$ results in a runaway of
the boundary. Just as in the case of the shock wave analyzed in \cite{DAM1}
a singularity develops which is asymptotic to the boundary where it
diverges and there is a resulting event horizon. Thus for the physical
situation at hand, the critical value of $\ep$ for which the boundary
starts running away and a black hole is formed is the lowest value of
$\ep$ for which any of the conditions (\ref{eq:thtwoa})-(\ref{eq:thtwob})
are satisfied, i.e. $\ep_{cr} = \aep_1$. When $\ep$ exceeds a higher
value $\bep_1$ the runaway occurs inside the
region of the incident pulse. The other runaways for higher values of
$\ep$ are irrelevant because they occur in a region which is already
behind the singularity and hence not contained in the semiclassical
space-time.
Profiles of the boundary for fixed value of $\alpha$ and different
$\ep$ (and hence different mass $M$) are shown in the Fig.1 - 3.

Note that we have kept $\alpha,\beta$ fixed in the above discussion.
It may be seen from (\ref{eq:thtwod}) that $\ep_{cr}$
decreases monotonically from $1$ at $\alpha = \alpha_0$ to
zero for infinite $\alpha$, with
$\ep_{cr} \rightarrow {\beta\sqd \over \alpha}$ as
$\alpha \rightarrow \infty$. Thus a limit of $\ep$
for some given $\alpha$ means that there is a lower limit on the
total mass of matter $M_{cr}(\alpha) = \alpha \ep_{cr}$
which has been thrown in.
For $\alpha = \alpha_0$ one has $M_{cr} = \alpha_0$. As $\alpha$ increases,
$M_{cr}(\alpha)$ first rises to a maximum value and then decreases,
eventually becoming asympotic to the constant shock wave value $\beta\sqd$
for large $\alpha$.

We thus conclude that for the boundary to run away and a black hole to
form there is a minimum value of the incoming energy density {\em as well}
as a minimum value of the total energy.

\section{Outlook}

The fact that the runaway of the boundary occurs at the same threshold
at which a black hole singularity is formed in the semiclassical model
is significant. While we have demonstrated this in the $N=24$ model, we
expect that a similar behaviour would hold for $N \neq 24$ as well. However
the differential euqtion in the latter case is higher order and the
stability analysis would be significantly different. It would be interesting
to get an idea of this by using numerical integration techniques. Some work
in this direction has been done in \cite{STHOR}.

For the shock wave solution we found in \cite{DAM1} that the runaway of
the boundary in the semiclassical theory also signals the threshold
beyond which an aypmtotic observer measures large quantum fluctuations
of the boundary near the horizon. It would be interesting to understand
the nature and implications
of the boundary flcutuations in a more general setting.
We leave that for future investigation.

\vskip .5cm
\noindent{\bf Acknowledgements:}
We would like to thank H. Verlinde for discussions.
{}~One of us (S.M.) would like to thank Professor Abdus Salam,
the International Atomic Energy and UNESCO for hospitality at the
International Centre for Theoretical Physics, Trieste. S.R.D. would
like to thank the Theoretical Physics Groups of Brown University,
Princeton University, Enrico Fermi Institute and Washington University
for hospitality during the final stages of this work.

\newpage
\begin{center}
FIGURES
\end{center}
\vskip 1cm

\noindent Fig.1: Typical boundary curve for $\alpha < \alpha_0$ for
any $\epsilon$. Even for $\alpha > \alpha_0$ and $M < M_{cr}(\alpha)$
the boundary curve is similar. No singularities are formed.

\vskip .5cm
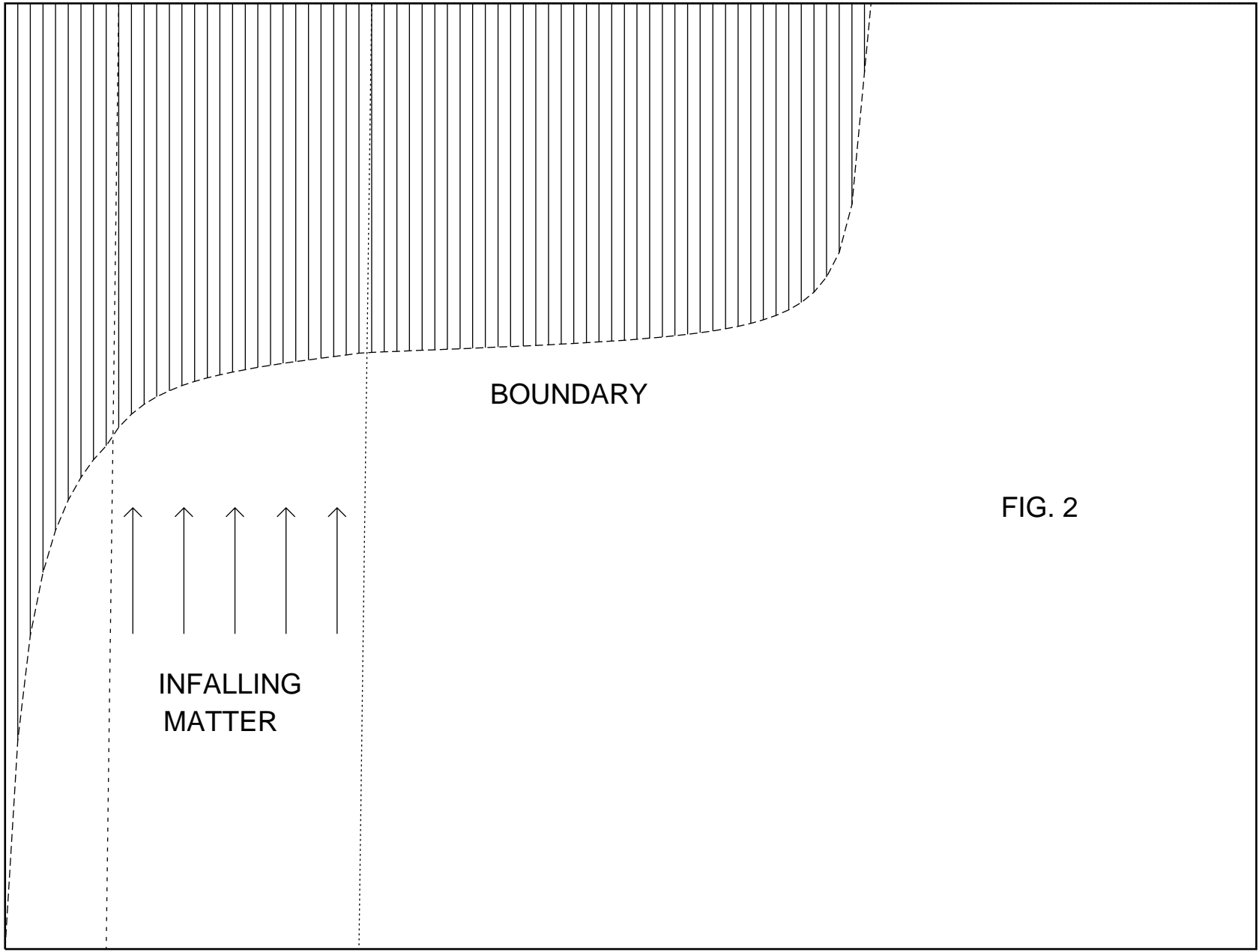
\noindent Fig. 2: Boundary curve for $\alpha > \alpha_0$ and $M
> M_{cr}(\alpha)$. Boundary runs away after the matter infall ends.
Cosequently space-time singularity is formed.
\vskip .5cm

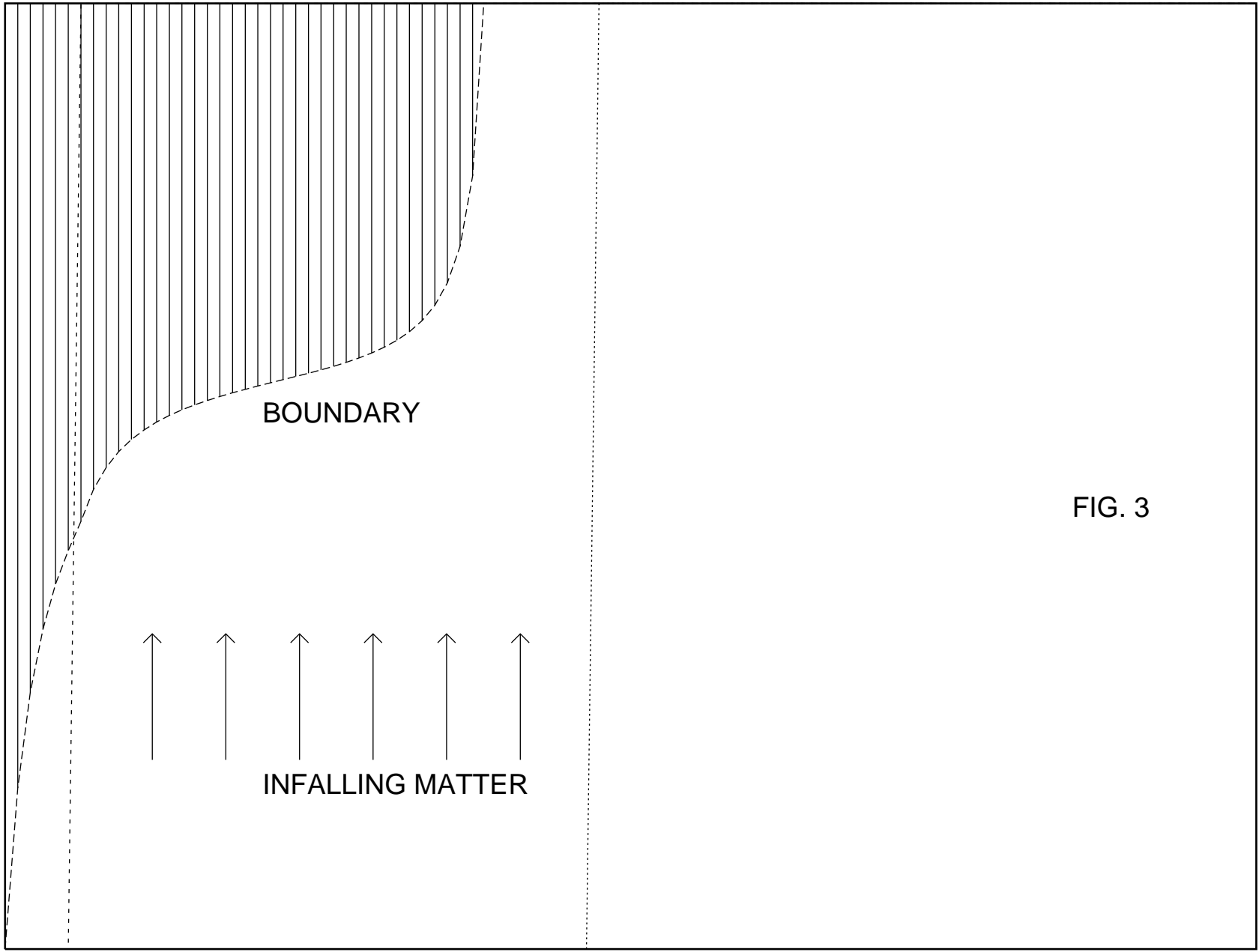
\noindent Fig. 3: For yet higher value of $M$, i.e. for the
corresponding $\ep > \ep_{cr}^{(2)}$, the boundary runaway
occurs in the region of incident pulse.
\end{document}